\documentclass[aps,preprint,preprintnumbers,amsmath,amssymb,nofootinbib]{revtex4}
\usepackage{amssymb}
\usepackage{lscape}
\usepackage{amsmath}
\usepackage{bm}
\usepackage{graphicx}
\usepackage{epsfig}
\begin{document}

\title{Pre-inflation: origin of the Universe from a topological phase transition}
\author{$^{1,2}$ Mauricio Bellini\footnote{E-mail address: mbellini@mdp.edu.ar} }
\address{$^1$ Departamento de F\'isica, Facultad de Ciencias Exactas y
Naturales, Universidad Nacional de Mar del Plata, Funes 3350, C.P.
7600, Mar del Plata, Argentina.\\
$^2$ Instituto de Investigaciones F\'{\i}sicas de Mar del Plata (IFIMAR), \\
Consejo Nacional de Investigaciones Cient\'ificas y T\'ecnicas
(CONICET), Mar del Plata, Argentina.}

\begin{abstract}
I study a model which describes the birth of the universe using a global topological phase transition with a complex manifold where the time, $\tau$, is considered as a complex variable. Before the big bang $\tau$ is a purely imaginary variable so that the space can be considered as Euclidean. The phase transition from a pre-inflation to inflation
is examined by studying the dynamical rotation of the time on the complex plane. Back-reaction effects are exactly calculated using Relativistic Quantum Geometry.
\end{abstract}
\maketitle

\section{Introduction}

One of the most important paradigm in modern cosmology is the
explanation of how the universe reached the inflationary epoch.
Inflation\cite{infl} is the more serious candidate to describe the
expansion of the universe since the Planckian time to about
$10^{-35}\,{\rm sec}$. This theory describes a quasi-exponential
expansion that can resolve the flatness, horizon and monopole
problems (among others). This theory has been very intensively
tested\cite{smoot} and provides a physical mechanism to explain
the generation of primordial energy density fluctuations on super
Hubble scales\cite{prd96}. The most conservative assumption is
that the energy density $\rho =P/\omega $ is due to a cosmological
parameter which is constant and the equation of state is given by
a constant $\omega =-1$, describing a vacuum dominated universe
with pressure $P$ and energy density $\rho$.

The theory that describes the earlier evolution of the universe is called pre-inflation\cite{Kur}.
The existence of a pre-inflationary epoch with fast-roll of the inflaton field would introduce an
infrared depression in the primordial power spectrum. This depression might have left an imprint in the CMB anisotropy\cite{GR}.
It is supposed that during pre-inflation the universe begun to expand from some Planckian-size initial volume to thereafter pass to an inflationary epoch.
Some models consider the possibility of an pre-inflationary epoch in which the universe is dominated by radiation\cite{IC}
In this framework Relativistic Quantum Geometry (RQG)\cite{rb}, should be very useful when we try to study the evolution of the geometrical
back-reaction effects given that we are dealing with Planckian energetic scales, and back-reaction
effects should be very intense at these scales\cite{ab}.

On the other hand, scalar fluctuations of the metric on
cosmological scales can be studied in a non-perturbative
formalism, describing not only small fluctuations, but the larger
ones \cite{re4}. However, the metric fluctuations can be studied
as a more profound phenomena in which the scalar metric
fluctuations appear as a geometric response to the scalar field
fluctuations by means of geometrical displacement from a Riemann
manifold to a Weylian one, through RQG.  The dynamics of the geometrical scalar field is defined on a Weyl-integrable manifold that preserves the gauge-invariance under the transformations of the Einstein's equations, that involves the cosmological constant. Our approach is different to quantum gravity.
The natural way to construct quantum gravity models is to apply quantum field theory methods to the theories of classical gravitational
fields interacting with matter. Our approach is different to quantum gravity because our subject of study is the dynamics of the geometrical quantum fields. This dynamics is obtained from the Einstein-Hilbert action, and not by using the standard effective action used in various models of quantum gravity\cite{od}. It is supposed that during pre-inflation the universe began to expand from some Planckian-size initial volume to thereafter pass to an inflationary epoch. In this framework RQG should be very useful when we try to study the evolution of the geometrical back-reaction effects given that we are dealing with Planckian energetic scales, and back-reaction effects should be very intense at these scales.

In this letter we explore the idea that the universe could be
created through a topological global phase transition, from a
Euclidean to an hyperbolic manifold, in which the time rotates in
the complex plane from the imaginary axis to the real one. When the
time reaches the real axis, the universe begins its inflationary
expansion. The back-reaction effects are studied in detail.

\section{The model}

With the aim to describe a big-bang theory we shall consider a complex manifold, in terms of which the universe describes a background semi-Riemannian expansion. The line element
for this case is
\begin{equation}\label{1}
d\hat{S}^2 = \hat{g}_{\mu\nu} d\hat{x}^{\mu} d\hat{x}^{\nu}= e^{2i\theta(t)} d\hat{t}^2 + a^2(t) \hat{\eta}_{ij} d\hat{x}^i d\hat{x}^j,
\end{equation}
with the signature: $(+,-,-,-)$. Here $\theta(t)=\frac{\pi}{2} \frac{a_0}{a}$, with $a\geq a_0$, $t$ is a real parameter time and $H_0=\pi/(2 a_0)=1/t_p$, such that $t_p=5.4 \times 10^{-44} \, {\rm sec}$ is the Planckian time. Notice that the metric (\ref{1})
describes a complex manifold such that, at $t=0$ the space-time is Euclidean, but after many Planckian times, when $\theta \rightarrow 0$, it becomes hyperbolic.
We shall define the background action ${\cal I}$ on this manifold, so that it describes the expansion driven by a scalar field, which is minimally coupled to gravity
\begin{equation}
{\cal I} = \int \, d^4x\, \sqrt{\hat{g}}\, \left[ \frac{{\cal \hat{R}}}{16 \pi G} +  \left[ \frac{1}{2}\dot\phi^2 - V(\phi)\right]\right],
\end{equation}
where $\sqrt{\hat{g}} = i a^3 e^{i\theta}$. After some algebraic work, we obtain that the relevant (complex) Einstein equations are
\begin{eqnarray}
3 H^2(t) e^{-i \hat{\theta}} & = & 8 \pi G \left( \frac{\dot\phi^2}{2} e^{-i\hat{\theta}} + e^{i\hat{\theta}} V(\phi)\right), \\
\left( 3 H^2 + 2 \dot{H} - 2i \dot{\hat{\theta}} H\right) e^{-i{\hat{\theta}}} & = & 8 \pi G \left( \frac{\dot\phi^2}{2} e^{-i{\hat{\theta}}} - e^{i{\hat{\theta}}} V(\phi)\right),
\end{eqnarray}
so that we can calculate the equation of state
\begin{equation}
\omega=P/\rho=-1 - \frac{2 \dot{H}}{3 H^2} + \frac{2i \dot{\hat{\theta}}}{3 H}.
\end{equation}
Here, the (complex) background pressure $P$ and the energy density $\rho$, are
\begin{eqnarray}
P & = & \frac{\dot\phi^2}{2} e^{-i\hat{\theta}} - e^{i\hat{\theta}} V(\phi), \\
\rho & = &\frac{\dot\phi^2}{2} e^{-i\hat{\theta}} + e^{i\hat{\theta}} V(\phi).
\end{eqnarray}
On the other hand, from the action (\ref{1}) we obtain
\begin{equation}\label{bac}
\ddot\phi + \left(3 \frac{\dot{a}}{a} - i\dot{\hat{\theta}}\right) \dot\phi + e^{2i\hat{\theta}} \frac{\delta{ V(\phi)}}{\delta\phi} =0,
\end{equation}
that describes the dynamics of the background field $\phi(t)$. The metric (\ref{1}) is not sufficiently explicit to describe the transition to an inflationary universe from a topological phase transition, because $t$ is not exactly
the dynamical coordinate that describes this transition. The correct dynamical variable in (\ref{1}) is: $\tau=\int e^{i \hat\theta(t)} dt$, which describes the time in the complex plane.
The idea is that $\tau$ becomes a space-like coordinate before the big bang, so that it can be considered as a reversal variable. However, after a phase transition we must require that it changes its signature and then can be considered as a causal variable. This dynamical change of signature describes a topological phase transition of the universe from an initial global Euclidean 4D space, to a final hyperbolic 4D space-time.

\section{An example: asymptotic de Sitter expansion}

We shall consider a scale factor, related to a de Sitter expansion in the $t$-dynamical scale: $H_0=\dot{a}/a(t)$, such that ${\cal H}(\tau)= \frac{1}{a(\tau)} \frac{d a(\tau)}{d\tau}=H_0\, e^{-i\hat{\theta}(\tau)}$, is
\begin{equation}\label{sf}
a(\tau) = a_0\, e^{{\rm Ei}\left[1, i\frac{\pi}{2} a_0 e^{-H_0 \tau}\right]}.
\end{equation}
Notice that we have used the fact that $\hat{\theta}(\tau) = \frac{\pi}{2} \frac{a_0}{a(\tau)}$. The expression (\ref{sf}) for the scale factor written in terms
of $\tau$ makes it very difficult to describe the cosmological dynamics of the universe. For this reason, we shall search for another variable to describe the dynamics of this cosmological phase
transition. A good candidate is the phase $\hat{\theta}$. Since $\hat\theta(t)=\frac{\pi}{2} e^{-H_0 t}$, we can rewrite the metric (\ref{1}), as
\begin{equation}\label{m}
d\hat{S}^2 =  \left(\frac{\pi a_0}{2}\right)^2 \frac{1}{\hat{\theta}^2} \left[{d\hat{\theta}}^2 + \hat{\eta}_{ij} d\hat{x}^i d\hat{x}^j\right].
\end{equation}
If we desire to describe an initially Euclidean 4D universe, that thereafter evolves to an asymptotic value $\hat{\theta} {\rightarrow } 0$, we must require that $\hat{\theta}$ to have an initial value $\hat\theta_0=\frac{\pi}{2}$. Furthermore, the nonzero components of the Einstein tensor, are
\begin{equation}
G_{00} = - \frac{3}{\hat{\theta}^2} , \qquad G_{ij} =  \frac{3}{\hat\theta^2 } \,\delta{ij},
\end{equation}
so that the radiation energy density and pressure, are respectively given in this representation by
\begin{equation}
\rho(\hat\theta) = \frac{1}{2\pi G} \frac{3}{(\pi a_0)^2}, \qquad\qquad
P(\hat\theta) = - \frac{1}{4\pi G} \frac{3}{(\pi a_0)^2}.
\end{equation}
The equation of state for the metric (\ref{m}), is
\begin{equation}
\omega(\hat\theta) =  - 1.
\end{equation}

We shall describe the case where the asymptotic evolution of the Universe is described by a vacuum expansion. In this case the asymptotic scale factor, Hubble parameter and the
potential are are respectively given by
\begin{equation}
a(t)= a_0\, e^{H_0 t}, \qquad \frac{\dot{a}}{a} = H_0 \qquad V= \frac{3}{8\pi G} H^2_0,
\end{equation}
so that the background field in (\ref{bac}), is
\begin{equation}
\phi(t)= \phi_0.
\end{equation}
This solution describes the background solution of the field that drives a phase transition of the global geometry
from a 4D Euclidean space to a 4D hyperbolic spacetime.

In order to describe the exact back-reaction effects, we shall consider Relativistic Quantum Geometry (RQG), introduced in
\cite{rb}. In this formalism the manifold is defined with a connection
\begin{equation}\label{gama}
\Gamma^{\alpha}_{\beta\gamma} = \left\{ \begin{array}{cc}  \alpha \, \\ \beta \, \gamma  \end{array} \right\}+ \sigma^{\alpha} \hat{g}_{\beta\gamma} ,
\end{equation}
such that the covariant derivative of the metric tensor in the Riemannian background manifold is null (we denote with a semicolon the Riemannian-covariant derivative): $\Delta g_{\alpha\beta}=g_{\alpha\beta;\gamma} \,dx^{\gamma}=0$, so that the Weylian covariant derivative\cite{weyl} on the manifold generated by (\ref{gama}) is nonzero:
$ g_{\alpha\beta|\gamma} = \sigma_{\gamma}\,g_{\alpha\beta}$. To simplify the notation we denote $\sigma_{\alpha} \equiv \sigma_{,\alpha}$. From the action's point of view, the scalar field $\sigma(x^{\alpha})$ is a generic geometrical transformation that leaves the action invariant\cite{rb}
\begin{equation}\label{aac}
{\cal I} = \int d^4 \hat{x}\, \sqrt{-\hat{g}}\, \left[\frac{\hat{R}}{2\kappa} + \hat{{\cal L}}\right] = \int d^4 \hat{x}\, \left[\sqrt{-\hat{g}} e^{-2\sigma}\right]\,
\left\{\left[\frac{\hat{R}}{2\kappa} + \hat{{\cal L}}\right]\,e^{2\sigma}\right\},
\end{equation}
Hence, Weylian quantities will be varied over these quantities in a semi-Riemannian manifold so that the dynamics of the system preserves the action: $\delta {\cal I} =0$, and we obtain
\begin{equation}
-\frac{\delta V}{V} = \frac{\delta \left[\frac{\hat{R}}{2\kappa} + \hat{{\cal L}}\right]}{\left[\frac{\hat{R}}{2\kappa} + \hat{{\cal L}}\right]}
= 2 \,\delta\sigma,
\end{equation}
where $\delta\sigma = \sigma_{\mu} dx^{\mu}$ is an exact differential and $V=\sqrt{-\hat{ g}}$ is the volume of the Riemannian manifold. Of course, all the variations are in the Weylian geometrical representation, and assure us gauge invariance because $\delta {\cal I} =0$.

\begin{equation}\label{de}
\frac{1}{\hat{\rho}} \frac{\delta \hat{\rho}}{\delta S} = - 2 \left(\frac{\pi}{2a_0}\right) \, \hat\theta \sigma',
\end{equation}
such that \cite{mb}, for ${\sigma}' = \left< (\sigma')^2 \right>^{1/2}$
\begin{equation}\label{fl}
\left< (\sigma')^2 \right> = \frac{1}{(2\pi)^{3}} \, \int d^3k ({\xi}_k)' \, ({\xi}^*_k)',
\end{equation}
where the modes $\xi_k$ must be restricted to
\begin{equation}
({\xi}_k^*)' \xi_k - ({\xi}_k)' \xi^*_k= i \hat{\theta}^2 \left(\frac{2}{\pi a_0}\right)^2,
\end{equation}
in order to the field $\sigma$ to be quantized\cite{rb}
\begin{equation}
\left[\sigma(x), \sigma_{\mu}(y)\right] =i \, \hbar \Theta_{\mu} \delta^{(4)}(x-y).
\end{equation}
Here, $\Theta_{\mu}=\left[\hat{\theta}^2 \left(\frac{2}{\pi a_0}\right)^2,0,0,0\right]$ are the components of the background relativistic tetra-vector on the Riemann manifold. The equation of motion for the modes of $\sigma$: $\xi_k(\hat\theta)$, is
\begin{equation}\label{mm}
\xi_k'' -   \frac{2}{\hat\theta} \xi'_k + k^2\, \xi_k(\hat\theta) =0,
\end{equation}
where the {\em prime} denotes the derivative with respect to $\hat\theta$.
The exact solution for the modes $\xi_k(\hat\theta)$ are
\begin{equation}
\xi_k(\hat\theta) = C_1(k) \left[ k\hat\theta \, \cos{(k\hat\theta)} - \sin{(k\hat\theta)}\right] +
C_2(k) \left[ k\hat\theta \, \sin{(k\hat\theta)} + \cos{(k\hat\theta)}\right].
\end{equation}
If we take $C_2(k)=i\,C_1(k)$, we obtain that:
\begin{equation}
C_1(k)= \frac{i}{2}\left(\frac{\pi}{2 a_0}\right) k^{-3/2}, \qquad  C_2(k)= -\frac{1}{2}\left(\frac{\pi}{2 a_0}\right) k^{-3/2},
\end{equation}
so that the quantized solution of (\ref{mm}) results to be
\begin{equation}
\xi_k(\theta)= \frac{i}{2}\left(\frac{\pi}{2 a_0}\right) k^{-3/2} \,e^{-i k \hat\theta} \left[k\hat\theta-i\right].
\end{equation}
Therefore, the fluctuations (\ref{fl}), are
\begin{equation}
\left< (\sigma')^2 \right> = \frac{1}{8} \frac{\hat\theta^2}{(4 a_0)^2} \epsilon^4 k^4_0,
\end{equation}
such that $\epsilon \ll 1$ and $k_0=\frac{\sqrt{2}}{\hat\theta}$. Hence, the amplitude of energy-density fluctuations on super Hubble scales, becomes
\begin{equation}
\left|\frac{1}{\hat{\rho}} \frac{\delta \hat{\rho}}{\delta S}\right| =  \frac{\pi \epsilon^2}{4\sqrt{2} a^2_0} ,
\end{equation}
which is a constant.

\section{Final Comments}

We have studied a model that describe the origin of the universe using a global topological phase transition from a
4D Euclidean manifold to an asymptotic 4D hyperbolic one. To develop this idea we have introduced a complex time, $\tau$.
The interesting of this idea is that $\tau$ is a space-like coordinated before the big bang, so that can be considered as a reversal variable. However, after the phase transition it changes its signature and then can be considered as a causal variable. Due to the fact that the description of the cosmological dynamics of the universe as a function of $\tau$ become complicated, we have expressed introduced the phase $\hat\theta$ as a dynamical variable.

As an example, we have explored the case where the universe evolves through an asymptotic de Sitter expansion.
We have studied back-reaction effects in a pre-inflationary universe using RQG. This formalism makes possible the non-perturbative treatment of the vacuum fluctuations of the spacetime,
by making a displacement from a semi-Riemannian to a Weylian one. In this description
the dynamics of the geometrical field $\sigma$ describes the geometrical quantum fluctuations with respect to the Riemannian (classical) background. In the example here studied,
we have found that back-reaction becomes frozen in time.

Finally, the field $\sigma$ has a geometrical origin, but must be interpreted as a primordial gravitational quantum potential, which generates a distortion of the metric. After pre-inflation, $\sigma$ could decay in different kinds of fields, one of which should be the inflaton field.

\section*{Acknowledgements}

\noindent  M. Bellini acknowledges
UNMdP and  CONICET (Argentina) for financial support.

\end{document}